\documentclass[aps, reprint,prl,superscriptaddress,nofootinbib]{revtex4-1}
\usepackage{dcolumn}
\usepackage[utf8]{inputenc}
\usepackage{amsmath}
\usepackage{amsfonts}
\usepackage{amssymb}
\usepackage{graphicx}
\usepackage{hyperref}
\usepackage{xcolor}

\usepackage{tikz}
\usetikzlibrary{quantikz}
\begin{document}

\title{Response to a comment on \\ ``Vindication of entanglement-based witnesses in hybrid quantum systems"}

\author{Chiara Marletto}
\affiliation{Clarendon Laboratory, University of Oxford, Parks Road, Oxford OX1 3PU, United Kingdom}
\affiliation{Centre for Quantum Technologies, National University of Singapore, 3 Science Drive 2, Singapore 117543, Singapore}

\author{Emanuele Marconato}
\affiliation{Dipartimento di Informatica, University of Pisa, Largo B. Pontecorvo 3, Pisa 56127, Italy}
\affiliation{Dipartimento di Ingegneria e Scienza dell'Informazione, University of Trento, Via Sommarive 9, Povo 38123, Italy}

\date{\today}%

\begin{abstract}
In a recent paper, \cite{MAMA}, we vindicated a general entanglement-based witness of non-classicality in hybrid quantum systems. Our vindication refutes a counterexample to the witness, proposed by Hall and Reginatto, \cite{HARE}. These authors recently commented further, \cite{HARE-pre}, claiming to expose ``a huge number of errors and misconceptions" in it. However, their comment contains no refutation of our arguments, nor does it expose any error or misconception in them. But it does include a number of misconceptions about the witness of non-classicality. Here we respond to those.

\end{abstract}

\maketitle

%\pacs{}

\section{Background}

{\bf Summary of the witness.} An experimental scheme for witnessing quantum effects in gravity is proposed in \cite{MAVE, SOUG}. This experimental scheme is underlain by a witness of non-classicality, \cite{MAVE, MRVPRD}. Informally, the witness says that assuming two physical principles (locality and interoperability of information, see \cite{MRVPRD} for the details), if a system $M$ can mediate ({\sl by local means}) entanglement between two quantum systems, $Q$ and $Q'$, then it must be non-classical. It is important to note that the phrase ``by local means" is used here to indicate a specific protocol, relying on assumptions that are outlined in \cite{SOUG, MAVE, MRVPRD, SOUG2}. For instance, $Q$ and $Q'$ must not interact with each other directly, but only via $M$. Also, the notion of non-classicality used here is defined via a general operational criterion in \cite{MRVPRD}. It is a generalisation of what in quantum theory is expressed {\sl formally} as ``having two distinct physical variables that do not commute" (see \cite{MRVPRD} for details). Informally, this means that those physical variables cannot be simultaneously measured to an arbitrarily high degree of accuracy by the same measuring apparatus.

{\bf Summary of the controversy.} In \cite{HARE}, Hall and Reginatto (HR) note ``that the claim" (of the witness) ``does rely on a specific model of quantum-classical dynamics, and hence does not provide an unambiguous witness of quantum mediators", adding that they seek to ``demonstrate there are counterexamples to the claim" (of the witness), ``based on the description of quantum-classical dynamics via the formalism of ensembles on configuration space ". Such counterexamples are intended to show that the experimental proposals based on the witness do not provide ``a definitive test of non-classical gravity''.

These claims are false, as we explain in the vindication, \cite{MAMA}. There, we stress that (as already discussed in \cite{MAVE, MRVPRD}) the witness does {\sl not} rely on a specific dynamical model (but on two general principles); and we offer a refutation of the proposed counterexample. HR's model is not a counterexample to the witness because: a) it conceals a hidden non-classicality of the mediator; and b) the hybrid ensemble model used in the counterexample violates one of the witness' assumptions (specifically, the principle of locality - which instead is satisfied by quantum theory). Therefore, observing gravitational entanglement between two quantum masses (generated {\sl by local means}, following the assumptions in \cite{SOUG, MAVE, MRVPRD}) is (contrary to what HR claim) a definitive test of non-classical gravity.

Recently HR commented further, \cite{HARE-pre}. The comment appears to be advocating a position that is inconsistent with their earlier paper \cite{HARE}. The latter presented a ``counterexample" to the proposals based on the witness of non-classicality.  But the comment \cite{HARE-pre} states instead that ``these models present no challenge to the mathematical validity of the theorem" (supporting the witness), ``and there is no a priori need for its vindication." This is inconsistent with the previously proposed model being a ``counterexample'',  \cite{HARE}. 

Moreover, HR display numerous misconceptions about the witness, about the notion of non-classicality, and about the vindication. Here we clarify only the most fundamental ones, also noting that their comment does not in fact address the criticisms proposed in \cite{MAMA}.  The key fallacy in HR's comment is an elementary misunderstanding of the implications of the witness. This leads them to address a position that is {\sl not} advocated by the vindication. For example, HR erroneously state that ``Marconato and Marletto [...] claim [...] that a theorem in Ref. \cite{MRVPRD} forbids the possibility of entanglement generation via a classical mediator for any model that satisfies two particular principles.'' HR have missed the point that the above statement applies to entanglement generation (under the said principles) occurring {\sl via a specific protocol}. This is why the statement holds true, contrary to HR's claim. 

Having missed that point, HR discuss in great detail a model with qubits (section 3.1, \cite{HARE-pre}) which violates the witness' assumptions and (hence) appears to be generating entanglement by a classical channel. This model is irrelevant. It is well-known that this and other models can lead to entanglement generation without resorting to a quantum mediator (Vedral and one of us already pointed this out in \cite{MAVE, MAVEQ}). All such models are not counterexamples to the witness because they violate one or more of the witness' assumptions. Specifically, they are not viable models for the experiment in \cite{MAVE, SOUG}. For example, some such models allow for interactions that the witness forbids (e.g. by resorting to direct interaction between the masses, controlled by a classical bit \cite{MAVE}); others violate fundamental physical principles (such as locality) that are satisfied by our best explanations of physical reality, and hence should be ruled out a priori. 

As already explained in \cite{SOUG, MAVE, MRVPRD}, the witness shows that there is an experimental scheme that can unambiguously rule out classical mediators (if entanglement is observed and some general principles are satisfied).  The experimental scheme has specific requirements on how the entanglement must be generated: one has to check that {\sl all} these requirements are satisfied (as one does with Bell's inequalities violation experiments, e.g. by independently testing each subsystem making sure that the tests are performed faster than it takes for light to travel between them). The witness then implies that if one observed entanglement between the two systems $Q$ and $Q'$ (generated via the protocol specified by the experimental scheme), then the models {\sl of that experiment} adopting a classical mediator and obeying the said two general principles would be experimentally ruled out.  

So the relevant dichotomy is {\sl not} (as HR say, \cite{HARE-pre}) between models ``which allow the generation of quantum entanglement via a classical mediator", and models which do not. Rather, the relevant dichotomy is between models which permit the generation of entanglement via a classical mediator {\sl by local means} (i.e. via the protocol outlined in \cite{SOUG, MAVE, MRVPRD}), under the two general principles of locality and interoperability, and those which do not.

As we explained in \cite{MAMA}, HR's proposed model falls within the latter class. 

{\bf Notation.} We briefly summarise the notation of the model proposed by HR as a ``counterexample" to the witness in \cite{HARE}. Two quantum systems $Q$ and $Q'$ interact with a third system, $C$ which is ``classical". Any point in the configuration space of the joint system is described by the triplet of real numbers $z=(q,q',x)$, representing each the position coordinate of $Q$,  $Q'$ and $C$. The map of classical and quantum observables onto the so-called {\sl ensemble observables} is defined as:
\begin{equation}\begin{aligned}
f(x,\nabla_x S) &\to C_f[P,S] := \int dz P f(x,\nabla_x S) \\
\hat M &\to Q_{\hat M}[P,S] := \langle \psi |\hat M |\psi \rangle \label{ENSOB}
\end{aligned}\end{equation}
for any classical observable $f$ and quantum observable $\hat M$. In both sectors, the {\sl ensemble observables} of the hybrid ensemble model, $C_f[P,S]$ and $Q_{\hat M}[P,S]$, are functionals of $(P(z),S(z))$, where the hybrid wavefunction is defined as, \cite{HARE}: 

$$\psi(q,q',x) = \sqrt{P(q,q',x)}\,e^{iS(q,q',x)/\hbar}\;.$$

\noindent One can introduce the hybrid Poisson brackets
$$ \{ A, B \}_H := \int dz  \Big [  \frac{\delta A}{\delta P} \frac{\delta B}{\delta S} - \frac{\delta A}{\delta S}\frac{\delta B}{\delta P} \Big ] $$
where $\frac{\delta A}{\delta f}$ is the variational derivative of the functional $A[f]$ with respect to $f$. This bracket induces a map from the algebra of the quantum and classical sectors to the Poisson algebra of ensemble variables $C_f$ and $Q_{\hat M}$ in the configuration space, defined as follows:
\begin{equation}\label{COMM}
\begin{aligned}
\{ C_f, C_g\}_H = C_{\{ f,g\}_P}\; , \; \{ Q_{\hat M}, Q_{\hat N}\}_H = Q_{[\hat M, \hat N]/i\hbar}
\end{aligned}
\end{equation}
with $f,g$ and $\hat M, \hat N$ being respectively a pair of generic classical and quantum observables. Via this representation, the state of the composite system is specified by the probability density $P(z)$ and its conjugate density $S(z)$, while the time evolution driven by the Hamiltonian $H$ reads
\begin{equation}
\frac{\partial P}{\partial t} = \frac{\delta H}{\delta S}\;,\;\frac{\partial S}{\partial t} = -\frac{\delta H}{\delta P} \;.\nonumber
\end{equation}

In the counterexample, the interaction Hamiltonian of the hybrid system $Q$, $Q'$ and $C$ is given by:
\begin{equation} \label{EqHam}
H[P,S]= g_1 \int dq dq' dx P(\partial_q S) x +g_2 \int dq dq' dx P(\partial_x S) q'\;. 
\end{equation}
This interaction couples each of $Q$ and $Q'$ with $C$ (with $g_1$ and $g_2$ being the respective coupling constants), while $Q$ and $Q'$ do not interact directly.

This Hamiltonian induces the following evolution (see \cite{HARE}):
\begin{eqnarray} \label{EqFinal}
\psi_t (q,q',x)= e^{ -it H_{eff} /\hbar}\psi_0(q,q',x),
\end{eqnarray}
where $H_{eff}=(g_1\hat p\hat x+g_2 \hat q' \hat k)$,  and $\hat p=-i\hbar \partial_q$ and $\hat k={-i\hbar}\partial_x$ can be regarded as the momenta operators of $Q$ and $C$ respectively. 
By starting the three system in say a product state $\psi_0(q,q',x)= \psi_Q(q)\psi_{Q'}(q')\psi_C(x)$, HR claim that it is possible (via a protocol involving the above interaction Hamiltonian) to obtain a state $\psi_t (q,q',x) $ where $Q$ and $Q'$ appear to be formally entangled. This is considered by HR a ``general counterexample''  to the witness of non-classicality.

\section{The vindication's two main criticisms}

In the vindication, \cite{MAMA}, we explained that the proposed counterexample does not invalidate the witness, using two arguments. Neither argument has been addressed in HR's recent comment, \cite{HARE-pre}, as we explain below. 

{\bf Non-classicality criticism.} In \cite{MAMA}, we noted that the model underpinning the HR's counterexample is inconsistent, as it conceals a hidden non-classicality. On the one hand, HR require the mediator $C$ to be classical -- i.e., all the observables of ${C}$ must be functions of $x$ and $\nabla_x S$ {\sl only}. In HR's own words (\cite{HARE}): ``the
configuration-ensemble dynamics correspond to a formal embedding of the classical particle C into a quantum system [...]. However, this correspondence does not extend to treating C itself as fully equivalent to a quantum system, since classical observables  cannot in general be represented by functions of $x$ and $\hat k$."
On the other hand, if entanglement is to be detected at the end of the protocol, then the non-classical variable of $C$ ($\hat k$, `complementary' to $x$) must be indirectly measurable, thus being an observable after all.  The reason why the variable $\hat k$ turns out to be physically observable is precisely that it participates in the dynamics which generates entanglement between the two quantum systems, defined by the Hamiltonian in \eqref{EqHam}. When confirming entanglement on the two systems $Q$ and $Q'$, at the end of the experiment, one in fact performs an indirect (`tomographic') measurement of $\hat k$. 

In other words, the protocol of generating entanglement between two systems, by local interaction via a mediator, is in fact an indirect measurement of the non-classical variable $\hat k$ of the mediator, \cite{MRVPRD}.  Hence it is incompatible with the hybrid ensemble model's central tenet that $\hat k$ is {\sl not} an observable and that the mediator is classical. 

This criticism has not been addressed in \cite{HARE-pre}. That is partly because HR misunderstand a key point: the reason why the variable $\hat k$ of the classical sector $C$ satisfies the operational criterion of non-classicality is {\sl not} (contrary to HR's claim) the formal property that $\hat k$ does not commute with $x$. It is that $\hat k$ is not measurable to arbitrarily high accuracy jointly with $x$. This requirement is a direct implication of the fact that $Q$ and $Q'$ are quantum systems, and that they must interact with $C$ via the interaction Hamiltonian \eqref{EqHam} (see\cite{MAMA, MRVPRD}). The property that $[x,\hat k]\neq 0$ is a formal ``shorthand" one can use to express that fact. But the physical meaning of this formal inequality is in the limitations on how accurately the two variables $x$ and $\hat k$ can be distinguished from one another - which is implied by the requirement that $Q$ and $Q'$ be quantum-mechanical systems, and that they interact with $C$ via \eqref{EqHam}. 

HR attempt a {\sl reductio ad absurdum} of the argument that $\hat k$ is a non-classical observable of the system $C$. They reason that if one assumed the two system $Q$ and $Q'$ to be classical, then our argument in \cite{MAMA} would likewise lead to the conclusion that the mediator $C$ must be non-classical in this case too. This claim is false. HR's erroneous argument is based on the fact that both in the quantum and the classical cases the {\sl formal} expression of the global state (after the interaction) is identical, as in \eqref{EqFinal}. But it overlooks the crucial fact that (despite the same formal appearance of the final state) the physical situations are radically different, due to their different counterfactual properties. When the two systems $Q$ and $Q'$ are classical, no entanglement is generated: in order to verify that two systems are entangled, it must be possible to violate Bell's inequalities (or satisfy equivalent operational criteria), which require the two systems to be quantum-mechanical in the first place. If $Q$ and $Q'$ are classical systems, that is not possible; hence in this case our argument would {\sl not} conclude  that $C$ is non-classical. A simple example is a correlated phase space distribution of two classical systems. The fact that the joint probability density is not equal to the product of the individual probability densities does not mean that the systems are entangled. They are simply classically correlated, meaning that the respective probabilities are not in general independent.

The fact that in both cases, formally, the variables $\hat k$ and $x$ do not commute is irrelevant. In the fully classical case, that non-commutativity is not associated with any physical constraint on the measurability of these two variables. Such constraints are present instead in the case where $Q$ and $Q'$ are quantum systems, as pointed out in \cite{MAMA, MRVPRD}. Therefore HR's criticism of the vindication is invalid: our reasoning, when correctly interpreted, does not lead to any absurd conclusion.

In summary, there are two cases: either one considers the situation already examined in the vindication, \cite{MAMA}, where the two systems $Q$ and $Q'$ are quantum-mechanical and there is entanglement at the end of the protocol -- in this case, the vindication's ``non-classicality" criticism applies; or one has the situation where $Q$ and $Q'$ are classical and no entanglement is generated overall -- which is perfectly compatible with the witness, in a trivial way: the mediator is classical and (hence) there is no entanglement.  

{\bf Non-locality criticism.} The other criticism we proposed in \cite{MAMA} is that HR's hybrid model violates the principle of locality (which, as we mentioned, is one of the two physical principles on which the witness relies). By `principle of locality' we mean the principle of no action at a distance (sometimes also called `microcausality condition' in quantum field theory \cite{HAAG}, or `Einstein's locality' \cite{PADE}). This is a stronger requirement compared with no-signalling, as the latter simply mandates that the locally empirically accessible variables of a given system cannot be affected by the action of transformations acting on space-like separated systems; the principle of no action at a distance instead requires that no local property can change, whether locally observable or not.  This principle is satisfied both by general relativity as well as by (relativistic and non-relativistic) quantum theory. 

As we discussed in \cite{MAMA}, in the case of general interactions the hybrid ensemble model violates no action at a distance, as it violates the property of ``strong separability" (i.e., $\{Q_{\hat M}, C_f\}=0$ for all classical and quantum observables). The violation occurs when considering entangled subsystems, exactly like in the specific model proposed by HR, where $Q$ and $Q'$ must become entangled. Moreover, with entangled systems, local dynamical variables (such as $\nabla_q S$) of one system can be modified by transformations on space-like separated systems, due to the non-separable nature of the hybrid wavefuction. 
In order to avoid a violation of strong separability, in \cite{HARE} HR impose what one may call a ``superselection rule", restricting the allowed interactions to be linear functionals of $\nabla_x S $ and $x$ (as in equation \eqref{EqHam}). As we say in \cite{MAMA}, the superselection rule is problematic as it has no analogue in standard quantum or classical physics, where instead more general interactions are allowed. 

HR do acknowledge that there is a violation of strong-separability (and hence of no action at a distance) in the general case of a hybrid ensemble model. But they dismiss the criticism, simply commenting that it is a `weakness' for the non-classicality witness to rely on the principle of no action at a distance. This line of argument is mistaken, because it neglects that the principle of no action at a distance is a pillar of both quantum theory (relativistic and non-relativistic, as evident in the Heisenberg picture) as well as of general relativity. So, far from being a `weakness', assuming this principle is a necessity to be faithful to our current best physical theories. In fact, the violation of no action at a distance in HR's hybrid ensemble model is sufficient to disqualify it as a faithful description of interacting quantum and classical systems, given that the latter both satisfy the principle of no action at a distance for any interactions (according to our best explanations of physical reality).
HR also comment that ``the fields are never gravitationally decoupled from the spacetime, due to the nature of the interaction, so
that there is always nonlocality''. This is false: in quantum field theory, even if fields are always coupled to charges, the principle of no action at a distance is fully satisfied, so there is no non-locality. 

Instead of addressing the non-locality criticism, HR simply restate what is well-known, i.e., that the hybrid ensemble model satisfies no-signalling when subject to special superselection rules. Hence this criticism, too, remains valid.

{\bf Generality of the criticisms.} In the comment, HR emphasise a particular `post-selection' step of the protocol in the counterexample, that utilises a measurement on the classical sector. In fact, this step makes the alleged counterexample even more problematic (as we discuss in the next section). Besides, the presence of this step is irrelevant, as it does not address the above two criticisms. This is because those criticisms point out general problems about the hybrid ensemble model and, in particular, about the interaction Hamiltonian \eqref{EqHam}, and hence apply to all protocols using that Hamiltonian to achieve entanglement.

\section{Comments on a qubit model}

In \cite{HARE-pre}, HR focus on a protocol which uses an intermediate measurement on the classical sector as a `post-selection'. They also provide a toy model with qubits to illustrate the idea.  The protocol with qubits is to prepare one of two orthogonal entangled states (say $\rho_1$ and $\rho_0$) of two qubits, {\sl conditioning} this preparation on a classical bit $c$ (thereby introducing correlations between the bit and the qubits). Then, upon revealing the value of $c$ by an appropriate measurement, the two qubit (initially in a mixed, separable state) can be locally rotated in one of the two prepared entangled states, $\rho_1$ or $\rho_0$ -- by use of a local operation (on either of the qubits) conditioned on the value of $c$. 

As mentioned in \cite{PAL2021}, for protocols of this kind to work, during the preparation one must generate correlations in the joint system of the mediator (the bit) and the probes (the qubits). The correlations are then redistributed (or localised) on the probes by a local measurement on the mediator. This phenomenon has long been known in information theory, \cite{CIRAC-ENT}, and it is a feature of LOCC protocols with quantum systems \cite{LOCC}. As acknowledged by HR, the assumptions of the witness rule out this scheme as an allowed means of generating entanglement between the probes: all the protocols discussed in \cite{MAVE, SOUG, MRVPRD} assume a product state to be prepared at the outset; or, as in \cite{MRVPRD}, a state where the relevant initial observables are sharp with a given value on all the three subsystems, so there must be no correlations pre-established between the classical sectors and the two quantum systems. Given that this particular example violates the assumptions of the protocols to which the witness applies, it cannot be used as a counterexample to the witness. Hence, appealing to this kind of `postselection' procedure makes the model proposed by HR even more problematic. 

As a side remark, the toy model assumes a preparation using entangling gates between the two systems $Q$ and $Q'$, conditioned on the value of the bit $c$. This conceals a circularity: one has then to explain how that entangling interaction between $Q$ and $Q'$ is realised in the first place. To do so, one invariably must go back to the scenario of having to quantise some mediating system, if a direct interaction between $Q$ and $Q'$ is disallowed (according to the assumptions of the witness) and the two said principles are still obeyed. 

A genuine (and interesting) counterexample to the witness would be a model where a classical mediator can entangle two quantum systems by local means, using the protocol envisaged in \cite{SOUG, MAVE, MRVPRD}, under the two said general principles. On the other hand, a protocol of the kind discussed by HR violates one or more of the witness' assumptions, and hence it is not a counterexample. As mentioned in the introduction, one can think of many such examples. For instance, \cite{MAVE}, it is well-known that a classical `control' bit can entangle $Q$ and $Q'$ by direct interaction. This is in no way a refutation of the witness, given that it achieves entanglement by a different protocol. In conclusion, contrary to what HR claim, this toy model does not invalidate our statement that “when observing entanglement  between the two masses, one can rule out all classical theories of gravity obeying the above-mentioned general principles”,\cite{MAMA}. This is because the statement refers (as it is clear given its context) to entanglement generated {\sl by local means}, via the protocol to which the witness applies. Hence, {\sl pace} HR, that statement still holds true.

\section{Clarifying a few other misconceptions}

In section 5 of their comment \cite{HARE-pre}, HR also discuss the assumptions of the theorem underlying the witness, (\cite{MRVPRD}). In doing so, they do not distinguish between what is a physical principle required for the witness to hold (there are two principles, as we mentioned: locality and interoperability of information); what is a simplifying assumption, that can be relaxed; and what is a feature of the protocol to which the witness applies. As a result, HR misunderstand both the witness and our vindication of it. We turn now to refuting these incorrect claims.

Point (iii) is erroneous: the witness does not assume that the interaction with the mediator have ``identical forms" on both $Q$ and $Q'$. HR must have misunderstood the assumption that the {\sl task} performed on each of them (a copy-like operation) is the same; in fact, any interaction that implements that task will do.  Moreover (as stressed in \cite{MAVE, MRVPRD}), the assumption that the mediator is a bit and the two systems being entangled are qubits (HR's points (i) and (ii)) is simply implemented for convenience of exposition, but the proof is directly generalisable to higher-dimensional systems with methods that are customary in information theory. For how discrete systems can approximate the continuum, we refer the reader to the well-known literature on the universality of quantum and classical computation, and on continuum-variable quantum computing. HR's points (iv) and (v) simply confirm that their proposed model is not a faithful model for the experiments (\cite{MAVE, SOUG}) to which the witness applies, and hence it is not a counterexample. As we explained, the post-selection procedure on the classical sector makes the proposed counterexample even more problematic as a way to generate entanglement by classical means under the witness' assumptions; moreover, it does not resolve the two main criticisms in our vindication. Also, (point (iv)) HR overlook that the very protocol discussed in their counterexample implies the possibility of preparing two (or more) distinguishable entangled states in output: how would any entanglement be testable in the laboratory, if only one state of $Q$ and $Q'$ could be prepared? 

Finally, HR comment on the notion of non-classicality in \cite{MRVPRD}, with a few erroneous remarks. The formal definition given in \cite{MRVPRD}, far from being ``counterintuitive" or ``circular" as HR claim, implies the informal property of having two distinct physical variables that cannot be simultaneously measured (in a single-shot fashion). It is a natural generalisation of the property of {quantum} complementarity, which is formally expressed in quantum theory by non-commutativity. There is no circularity in the proposed definition: the statement quoted by HR (about a non-classical mediator enabling ``non-classical tasks on other superinformation media, such as establishing entanglement" \cite{MRVPRD}) is not a definition, but a commentary on the definition of non-classicality, to illustrate its meaning. On the other hand, what HR quote as their proposed definition of classicality (`` in our definition of classicality the observables of a classical system are in one-to-one correspondence with functions of position and momentum on a classical phase space", \cite{HARE-pre}) is circular and based on formal, non-physical concepts. This circularity is a manifestation of a general problem afflicting theories of classical information that do not model the task of `distinguishing' physically. In \cite{DEMA} Deutsch and one of us have solved this problem, in an elegant, non-circular way, where distinguishability is defined in terms of `copiability'. 

\section{Conclusion}

We have scotched a few misconceptions about the witness of non-classicality in support of recently proposed tests of non-classicality in gravity, \cite{SOUG, MAVE, MRVPRD}, also correcting erroneous claims made by HR in regard to our vindication, \cite{MAMA}, and about the witness itself.  The vindication has not been refuted: HR's proposed model is not a valid counterexample to the witness, because it has several difficulties. The two most serious ones are that: 1) it adopts a notion of `observables' of the classical sector that is inconsistent with the claimed ability of that sector to generate entanglement and 2) it is ruled out by the principle of no action at a distance.  Adding post-selection procedures, which anyway deviate from the prescription of the protocol to which the witness applies, does not solve these two issues. In fact, the post-selected protocol suffers from additional problems, such as relying on entanglement-generating mechanisms that in turn are not explained by the HR hybrid ensemble model.  

Important implications of this discussion are: one, that the experimental proposals supported by the witness are valid tests of non-classicality, based on general principles; two, that the notions of `observable' and `non-classicality' must be defined physically, not formally -- to do that, one needs to adopt a more general take on information theory, \cite{DEMA}, based on the notion of `copiability'; and three (as already remarked by many authors (see e.g. \cite{SAL, TER})) that violating the universality of quantum theory in order to fit quantum phenomena within a semi-classical picture of the world causes serious difficulties. We wonder how much longer it will take before physicists finally take quantum theory seriously. \\ {\bf Acknowledgements} \;\;The authors thank D. Deutsch, L. L. Salcedo, and V. Vedral for useful comments. This research was made possible through the generous support of the Eutopia Foundation and of the ID 61466 grant from the John Templeton Foundation, as part of the The Quantum Information Structure of Spacetime (QISS) Project (qiss.fr). The opinions expressed in this publication are those of the authors and do not necessarily reflect the views of the John Templeton Foundation.

\end{document}